\documentclass[twocolumn,letterpaper,floatfix]{revtex4}

\usepackage[english]{babel}
\usepackage{graphicx}
\usepackage{amssymb}

\begin{document}

\title{Quantitative analysis of the dripping and jetting regimes in co-flowing capillary jets} 

\author{Mar\'ia Luisa Cordero$^1$, Fran\c{c}ois Gallaire$^2$ and Charles N. Baroud$^1$}

\affiliation{$^1$LadHyX and Department of Mechanics, Ecole Polytechnique, CNRS, 91128 Palaiseau, France\\
$^{2}$ Laboratory of Fluid Mechanics and Instabilities (LFMI), Ecole Polytechnique F\'ed\'erale de Lausanne (EPFL) EPFL/STI/IGM/LFMI, CH-1015 Lausanne, Switzerland}

\begin{abstract} We study a liquid jet that breaks up into drops in an external co-flowing liquid inside a confining microfluidic geometry. The jet breakup can occur right after the nozzle in a phenomenon named dripping or through the generation of a liquid jet that breaks up a long distance from the nozzle, which is called jetting. Traditionally, these two regimes have been considered to reflect the existence of two kinds of spatiotemporal instabilities of a fluid jet, the dripping regime corresponding to an absolutely unstable jet and the jetting regime to a convectively unstable jet. Here, we present quantitative measurements of the dripping and jetting regimes, both in an unforced and a forced state, and compare these measurements with recent theoretical studies of spatiotemporal instability of a confined liquid jet in a co-flowing liquid. In the unforced state, the frequency of oscillation and breakup of the liquid jet is measured and compared to the theoretical predictions. The dominant frequency of the jet oscillations as a function of the inner flow rate agrees qualitatively with the theoretical predictions in the jetting regime but not in the dripping regime. In the forced state, achieved with periodic laser heating, the dripping regime is found to be insensitive to the perturbation and the frequency of drop formation remains unaltered. The jetting regime, on the contrary, amplifies the externally imposed frequency, which translates into the formation of drops at the frequency imposed by the external forcing. In conclusion, the dripping and jetting regimes are found to exhibit the main features of absolutely and convectively unstable flows respectively, but the frequency selection in the dripping regime is not ruled by the absolute frequency predicted by the stability analysis.
\end{abstract}

\maketitle

\section{Introduction}

A liquid column that exits from a nozzle in the form of a jet tends to break up into droplets due to surface tension. During the breakup process other forces are also present and oppose the action of surface tension. In particular, viscous forces damp the growth of the varicose deformations of the jet that lead to the pinch off and inertial forces promote the formation of a long liquid thread. Gravitational effects can also oppose the action of surface tension by elongating a falling liquid thread. These forces define the parametric space which determines the breakup dynamics of the liquid jet.

\begin{figure*}[t]
\includegraphics[width=17cm]{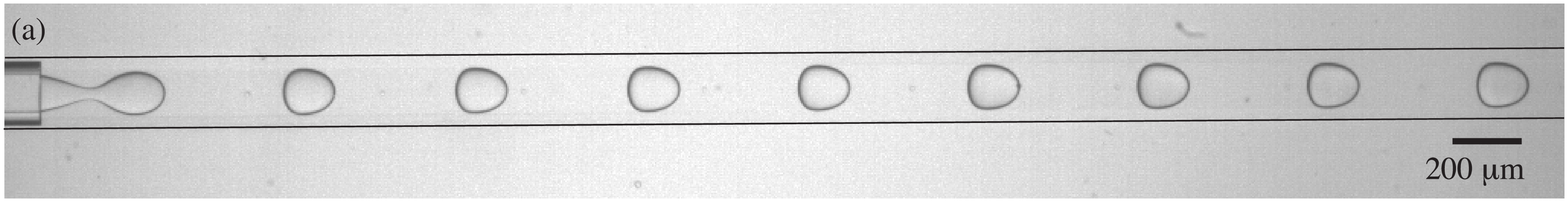}\\
\includegraphics[width=17cm]{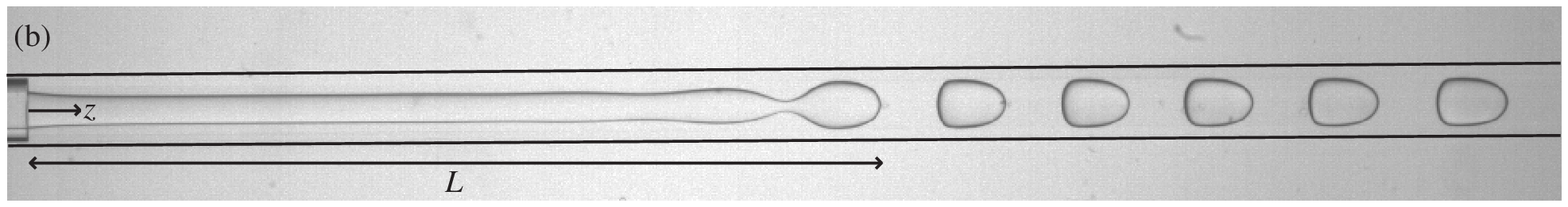}
\caption{\label{fig:Dripping-Jetting}Snapshots of the system in the dripping (a) and in the jetting (b) regimes. Experimental conditions are (a) $Q_i = 5 \ \mu$L/min, $Q_e = 30 \ \mu$L/min and (b) $Q_i = 14 \ \mu$L/min, $Q_e = 30 \ \mu$L/min.}
\end{figure*}

Liquid jets exhibit a large range of breaking patterns in different regions of the parameter space~\cite{Eggers2008}. As a general rule, the system exhibits a regime of periodic drop emission near the nozzle when capillary forces are dominant over other forces. This phenomenon is named dripping [Fig.~\ref{fig:Dripping-Jetting}(a)] and produces monodisperse drops whose size is typically determined by the nozzle size. As inertial or viscous forces become comparable to capillary forces the system transitions to a regime characterized by the formation of a long jet. Its radius undulates and it breaks up into drops far downstream from the nozzle. This regime is known as jetting [Fig.~\ref{fig:Dripping-Jetting}(b)] and usually exhibits a high dispersion in the drop size.

\begin{figure}[h]
\includegraphics[width=8.6cm]{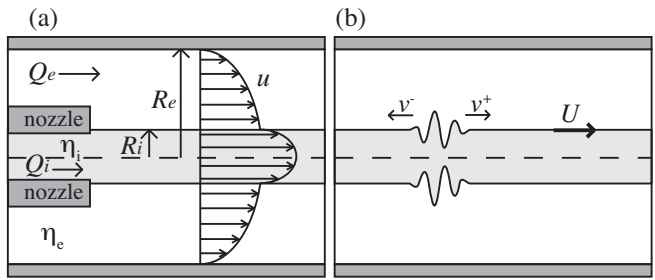}
\caption{\label{fig:Jets}(a) Co-flowing geometry. Inner fluid has viscosity $\eta_i$ and is injected at flow rate $Q_i$, while outer fluid has viscosity $\eta_e$ and is pushed at flow rate $Q_e$. (b) Schematic picture of the propagation of perturbations in a jet whose interface moves with velocity $U$. $v^-$ and $v^+$ represent the edge velocity of a wave packet composed of unstable perturbations.}
\end{figure}

The transition from dripping to jetting in a jet that falls in a quiescent outer fluid under the effect of gravity has been extensively studied, for example as a model of chaotic dynamics~\cite{Clanet1999, Ambravaneswaran2000, Coullet2005, Subramani2006, LeDizes1997}. On the scale of microfluidics, early studies of the transition were performed by Ga\~n\'an-Calvo~\cite{Ganan-Calvo1998}. In comparison, the dripping to jetting transition in confined liquid jets that flow in an immiscible carrier fluid has received less attention but interest in this geometry has increased since the recent development of microfluidic devices for the generation of emulsions~\cite{Baroud2010}. One of these devices is the so-called co-flowing geometry~\cite{Utada2007a}, in which the inner fluid is injected into a carrier fluid that is flowing in parallel [Fig~\ref{fig:Jets}(a)]. Gravitational effects are typically negligible in these systems. 

Recently, Utada et al.~\cite{Utada2007a} have studied this system when the transversal dimensions of the container channel is large compared to the jet radius and they have observed a dripping to jetting transition when viscous or inertial forces become comparable to capillary forces. This study was followed by several that considered the case of a highly confined jet with negligible inertia, which we concentrate on below. As shown by Guillot et al.~\cite{Guillot2007, Guillot2008}, the transition in this case is ruled by the capillary number Ca, which measures the ratio of viscous to capillary forces and whose critical value depends on the value of the confinement.

The theoretical efforts to understand the dripping to jetting transition in these devices have mainly concentrated on the linear stability of the liquid jet. Indeed, a liquid jet is unstable to axisymmetric perturbations of its base columnar state~\cite{Rayleigh1879a}. In the case of a flowing jet, perturbations can be modeled as exponential functions with complex frequencies and wavelengths. These perturbations can grow both in time and space while traveling along the jet interface [Fig.~\ref{fig:Jets}(b)]. If the growing perturbations travel both downstream and upstream, the instability is said to be absolute. In analogy with the high Reynolds number jets~\cite{LeDizes1997, Leib1986b, Monkewitz1990}, this case has been related to the dripping regime, since instabilities invade the whole jet and particularly the exit of the injection nozzle. Conversely, if all growing perturbations travel downstream the instability is said to be convective and this has been related to the jetting regime, in which the fluid velocity is high enough to advect the instabilities and induce breakup away from the nozzle.

Guillot et al.~\cite{Guillot2007, Guillot2008} and Herrada et al.~\cite{Herrada2008} have performed stability analyses for the confined liquid jet in a co-flowing liquid. They predict a transition from an absolute to a convective instability at a critical value of Ca, in the limit of negligible inertial forces. These authors find a good agreement between this theoretical critical value with the observed value of Ca at which the system switches from dripping to jetting, confirming that the two transitions are related. However, there exist no detailed comparisons between the theoretical predictions and the experiments beyond the threshold of the transition, which is the subject of this paper. In particular, we perform a frequency and wavelength analysis in both the jetting and dripping regimes and compare the measurements with the theoretical predictions. We also demonstrate the possibility to force the breakup of the liquid jet with an external stimulus, which allows us to control the drop size in the jetting regime.

In the next section we present the setup used for the experiments. Section~\ref{sec:Results} presents the experimental results. In Section~\ref{sec:BaseFlow} we present the measurements of the flow in the absence of external forcing. Then, in Section~\ref{sec:PerturbedFlow} we study the response of the flow to an external perturbation induced with laser heating. In Section~\ref{sec:Comparison} our experimental results are compared with the predictions of existent theories concerning confined liquid jets. Finally, Section~\ref{sec:Conclusions} summarizes the results and conclusions.

\section{Methods} \label{sec:Setup}

\begin{figure}[h] \centering
\includegraphics[width=8.6cm]{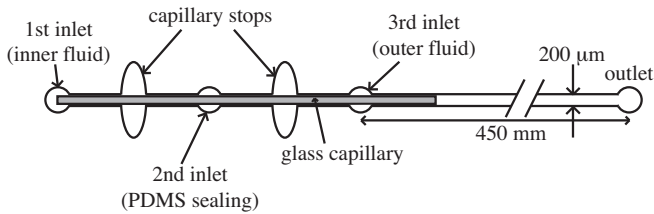}
\caption{\label{fig:Setup} Schematic representation of the channel geometry (not to scale).}
\end{figure}

Our experiments are performed inside a straight 5~cm long microchannel of square cross section. The microchannel mold is produced by lithography using four layers of a photoresist of thickness 50~$\mu$m each (Etertec XP-800-20) on a glass slide. The channel is then replicated in PDMS prepared in a 10:1.5 ratio of PDMS base with curing agent. Three inlets and one outlet are drilled in the PDMS block (see Fig.~\ref{fig:Setup}). A glass cylindrical capillary of external diameter 200~$\mu$m and internal diameter 100~$\mu$m is placed between the first and the third inlets to form the co-flowing geometry where the drops and jets are produced. The alignment of the glass capillary inside the square channel is ensured by the match between the size of the channel and the external diameter of the glass capillary. After bonding the channel to a glass slide, a drop of PDMS, prepared in a 10:0.5 ratio of PDMS base with curing agent, is allowed to enter the channel through the second inlet. After flowing by capillarity through the space between the PDMS walls and the glass capillary in both directions, the PDMS reaches a wider section of the channel where it stops flowing due to the change of Laplace pressure~\cite{Ody2007}. The channel is then baked to cure the PDMS and ensure the sealing between the glass capillary to the external channel. In this way, the inner fluid is forced to flow inside the glass capillary.

Fluids are injected at constant flow rates using syringe pumps (Kd Scientific 101, Harvard Apparatus PHD2000W). The inner fluid is a 80\% glycerol and 20\% water mixture, whose viscosity and density at $20^\circ$C~ are $\eta_i = 60.1$~mPa~s and $\rho_i = 1208.5$~kg/m$^3$ respectively. The outer fluid is silicone oil (Rhodorsil 47 V 100) of viscosity $\eta_e = 100$~mPa~s and density $\rho_e = 965$~kg/m$^3$. Surface tension between both fluids is measured as a function of temperature at $\gamma(T) = 47 - 0.3 \ T$~mN/m, where $T$ is the temperature, in degrees Celsius.

The channel is mounted on an inverted microscope (Nikon ECLIPSE TE2000-U) and observed through the microscope objective (Nikon, 4x/0.1 NA). Image sequences are recorded at 1000~frames per second with a fast camera (Photron Fastcam1024 PCI). An infrared laser of wavelength 1480~nm (Fitel Furukawa FOL1425) is focused inside the channel through the same objective, a distance 215~$\mu$m from the nozzle. The power of the laser beam is modulated with a signal generator (Wavetek model 80) to a sinusoidal signal of amplitude $P$ up to 55~mW and frequency $f_\textrm{\small laser}$ ranging between 10 and 150~Hz.

\begin{figure}[h] \centering
\includegraphics{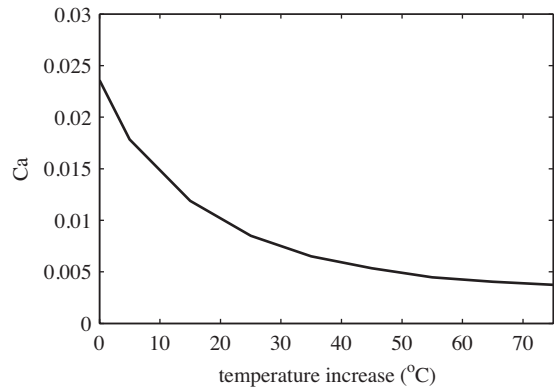}
\caption{\label{fig:Capillary} Capillary number as a function of temperature increase above room temperature for $U_m = 20$~mm/s.}
\end{figure}

The laser radiation is used to heat the inner fluid by absorption. This produces a periodic forcing of the flow by locally changing the capillary number Ca~$= \eta_i U_m/\gamma$ (where $U_m$ is the mean velocity of the flow), which controls the stability of the jet. Indeed, the temperature increase reduces both the interfacial  tension between the fluids $\gamma$ and the inner viscosity $\eta_i$~\cite{Handbook}. Figure~\ref{fig:Capillary} shows the effect of the temperature increase above room temperature ($25^\circ$C) on Ca for a typical value of the mean flow velocity $U_m = 20$~mm/s. We find that Ca decreases when the temperature increases, which means the laser acts by locally destabilizing the flow.

The temperature increase depends on the laser power. For the powers used here we expect the heating to range between $5^\circ$C and $30^\circ$C, based on our previous results~\cite{Cordero2009a}, which corresponds to a local decrease of Ca between 25\% and 70\% approximately. Compared to our previous work~\cite{Baroud2007a}, the laser heating here is weaker and the flow velocity and fluid viscosity larger, so the laser produces a small perturbation that cannot block the flow.

\section{Experimental results} \label{sec:Results}

\subsection{Unforced state} \label{sec:BaseFlow}

First, we study the flow behavior in the absence of the laser heating. The external flow rate is fixed at $Q_e = 30 \ \mu$L/min, while the internal flow rate $Q_i$ is increased from 5~$\mu$L/min to 32~$\mu$L/min with a step of 1~$\mu$L/min. A transition from dripping [Fig.~\ref{fig:Dripping-Jetting}(a)] to jetting [Fig.~\ref{fig:Dripping-Jetting}(b)] is observed at $Q_i = 14 \ \mu$L/min, similar to what was described elsewhere~\cite{Cubaud2008, Guillot2008}.

\begin{figure}[h]
\centering
\includegraphics{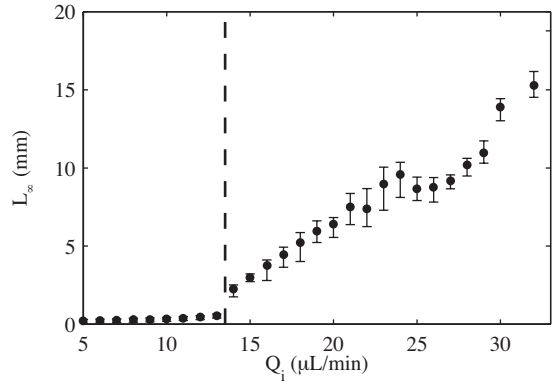}
\caption{\label{fig:LJet} Length of the inner fluid column as a function of the inner flow rate for an external flow rate fixed at $Q_e = 30 \ \mu$L/min.}
\end{figure}

The length of the inner fluid column, $L_\infty$, is measured as a function of $Q_i$ and is shown in Fig.~\ref{fig:LJet}. The error bars are calculated by considering the minimum and maximum values measured for $L_\infty$ during the drop formation process. For low $Q_i$, $L_\infty$ is small and grows only slightly with $Q_i$. This corresponds to the dripping regime. At $Q_i = 14 \ \mu$L/min, the jet starts to grow, showing that the jetting regime has been attained.

The low dispersion in $L_\infty$ in the dripping regime shows that drop breakup occurs at a steady position, suggesting that it is a regular, periodic process. This contrasts with the high dispersion in the jetting regime, which reflects the fact that the position of drop pinch off is not as well defined and therefore the drop formation is not regular in time.

To confirm this, we analyze the frequency of drop production in both regimes by measuring the frequency of the oscillations of the inner fluid column. The thickness of the jet is measured as a function of space and time from image analysis. This yields the radius of the jet $r(z,t)$, where $z$ is the axial direction shown in Fig.~\ref{fig:Dripping-Jetting}(b) and $t$ is time, for each value of $Q_i$. A discrete temporal Fourier transform yields the frequency spectrum of the oscillations of the jet's radius at each point $z$ of the jet. These spectra are averaged in space to obtain one spectrum for each value of the inner flow rate.

\begin{figure*}[t]
\centering
\includegraphics{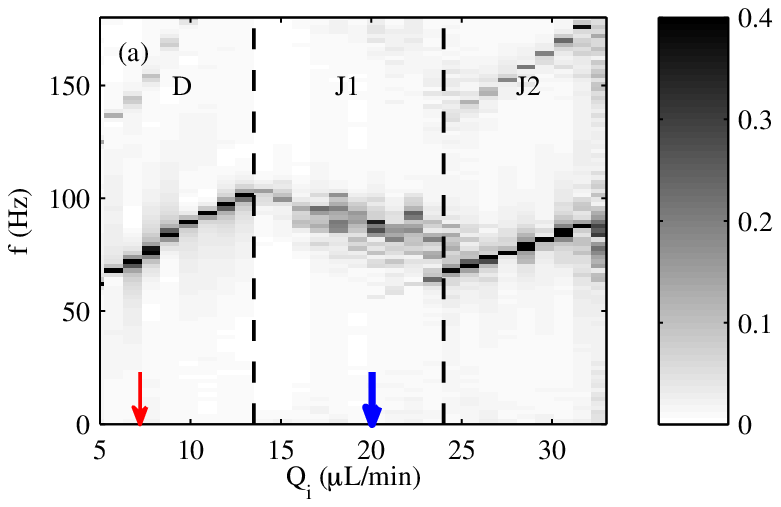} \includegraphics{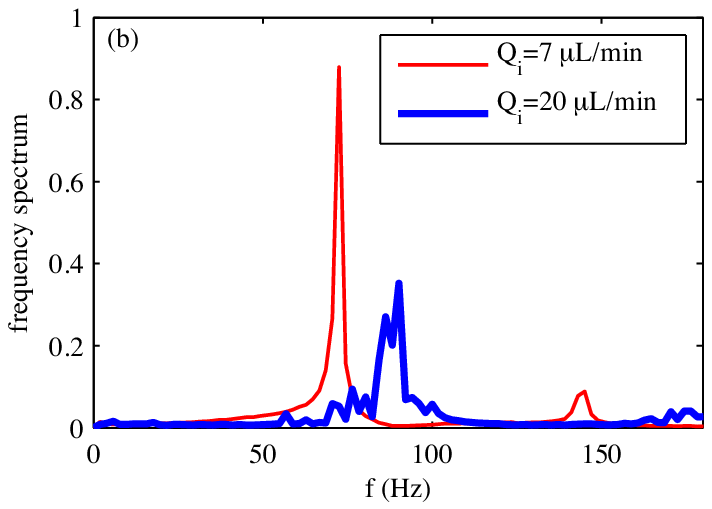}
\includegraphics{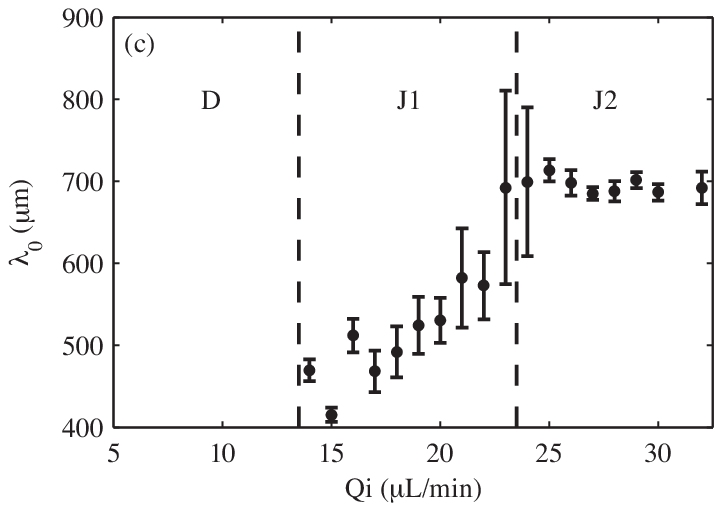} \includegraphics{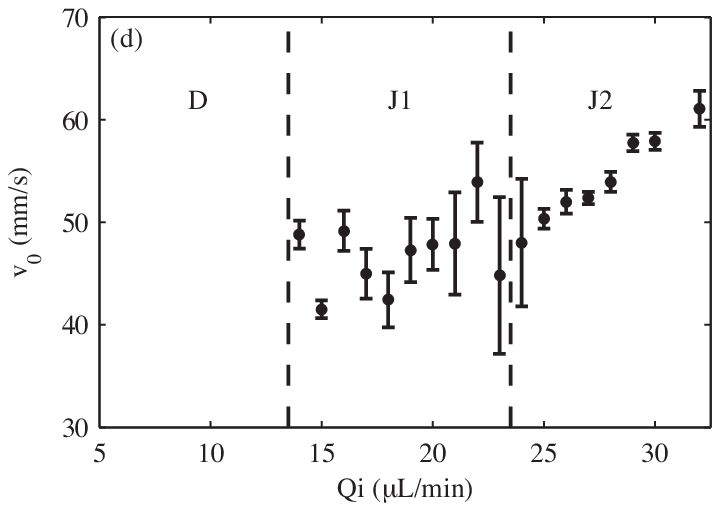}
\caption{\label{fig:BaseState} (Color online) (a) Frequency spectrum (in arbitrary units) of the jet oscillations as a function of $Q_i$. The red thin and blue thick arrows mark the flow rates $Q_i = 7 \ \mu$L/min and $Q_i = 20 \ \mu$L/min, which correspond to the experimental conditions of Figs.~\ref{fig:FFT}(a) and (b) respectively. The frequency spectra for these two particular cases are shown in (b) and represent a typical frequency spectrum in the dripping regime (red thin line) and in the jetting regime (blue thick line). (c) and (d): Wavelength and phase velocity of the jet oscillations respectively as a function of $Q_i$. In (a), (c) and (d), dashed  vertical lines are used to separate the different behaviors: Dripping (D), first jetting regime (J1) and second jetting regime (J2). In all cases $Q_e = 30 \ \mu$L/min.}
\end{figure*}

The frequency spectra as a function of $Q_i$ are shown in Fig.~\ref{fig:BaseState}(a) in gray scale. In the dripping region ($Q_i < 14 \ \mu$L/min) a sharp peak is observed, whose position increases linearly with $Q_i$. A typical spectrum for this region is shown in Fig.~\ref{fig:BaseState}(b). In the jetting regime, instead, two regions can be distinguished. First, for $14 < Q_i < 24 \ \mu$L/min, a broadening of the peak can be observed [see Fig.~\ref{fig:BaseState}(b)] with a global decrease in its position. We call this region ``first jetting regime'' or J1. Then, for $Q_i > 24 \ \mu$L/min, a sharp peak whose position increases linearly with $Q_i$ can be observed again. This region is called ``second jetting regime'' or J2. A close examination shows that the peak of J2 is associated with fluctuations induced by the injection of the inner fluid by the syringe pump (see Section~\ref{sec:Comparison}).

In contrast to the dripping regime, the jet in the jetting regime undergoes undulations characterized by a spatial wavelength visible on Fig.~\ref{fig:Dripping-Jetting}(b). The wavelength of these oscillations, $\lambda_0$, could not be accurately obtained from discrete spatial Fourier transform due to the insufficient spatial resolution of the optical setup. Instead, it was deduced from the drop volume $V_d$ and the mean jet radius $R_i$, calculated as the temporal and spatial average of $r(z,t)$.  Assuming volume conservation, we write

\begin{equation}
\label{eq:DropSize} V_d = \pi R_i^2 \lambda_0,
\end{equation}

\noindent which allows us to calculate $\lambda_0$. Figure~\ref{fig:BaseState}(c) shows $\lambda_0$ as a function of the inner flow rate. The error bars are calculated from the dispersion of the drop size distribution. In the first jetting regime, $\lambda_0$ increases with $Q_i$ and dispersion is large, reflecting the polydispersity of the drops. In the second jetting regime, $\lambda_0$ remains constant, with little dispersion.

Finally, the phase velocity of the jet oscillations is calculated from the frequency and wavelength of oscillations as $v_0=\lambda_0 f_0^\textrm{\small exp}$, where $f_0^\textrm{\small exp}$ is the position of the maximum of the frequency spectrum for each $Q_i$. As seen on Fig.~\ref{fig:BaseState}(d), $v_0$ remains constant at around 45 mm/s with large dispersion for the first jetting regime, due to the large error bars of $\lambda_0$. In the second jetting regime, $v_0$ increases linearly, with low dispersion.

\subsection{Forced state} \label{sec:PerturbedFlow}

\begin{figure}[h]
\centering
\includegraphics{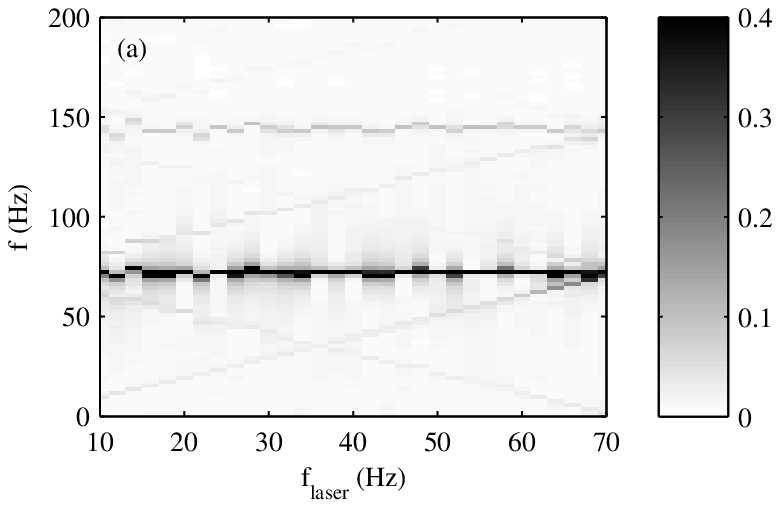}
\includegraphics{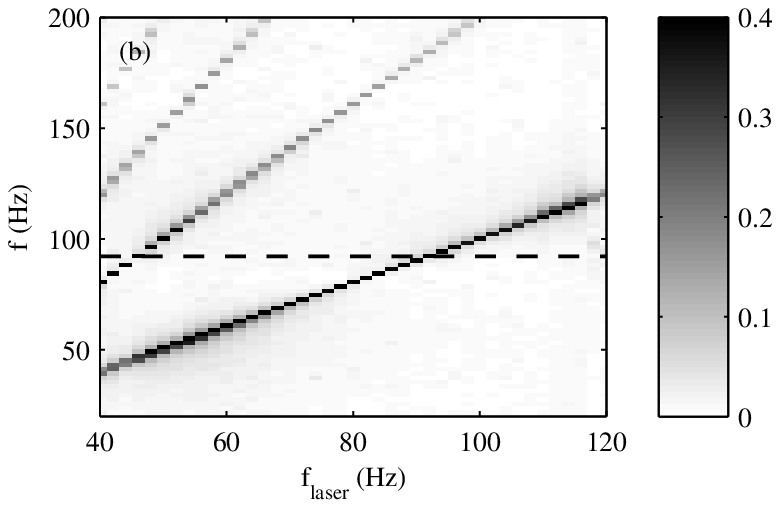}
\caption{\label{fig:FFT} Frequency spectrum of the jet oscillations as a function of $f_\textrm{\small laser}$ for (a) a dripping flow ($Q_e = 30 \ \mu$L/min, $Q_i = 7 \ \mu$L/min) and (b) a jetting flow ($Q_e = 30 \ \mu$L/min, $Q_i = 20 \ \mu$L/min) forced with laser heating. In both cases $P = 55$~mW. The dashed line in (b) shows the position of the
dominant frequency $f_0^\textrm{\small exp}$ in the unforced case.}
\end{figure}

Next, we study the response of the flow to an external periodic perturbation driven by the laser heating. For that, $Q_i$ and $Q_e$ are kept constant while the laser is modulated at different frequencies and powers. With the flow in the dripping regime ($Q_e = 30 \ \mu$L/min and $Q_i = 7 \ \mu$L/min), no noticeable effect is detected, even at the highest laser power. Frequency spectra of the jet oscillations are calculated for each value of the laser frequency $f_\textrm{\small laser}$, as explained above, and are shown in Fig.~\ref{fig:FFT}(a) for a laser power of 55~mW. These spectra show the dominance of the frequency peak at $f_0^\textrm{\small exp}$ which corresponds to the unforced case. Only at high laser power does the peak corresponding to $f_\textrm{\small laser}$ appear but its amplitude is much smaller than the dominant one, thus not changing noticeably the drop generation.

The same experiment is performed with the flow in the first jetting regime ($Q_e = 30 \ \mu$L/min, $Q_i = 20 \ \mu$L/min). For low laser power, no response of the jet is observed. When the laser power is increased, however, the jet's oscillations synchronize to the laser frequency for some frequency ranges. This can be observed in the frequency spectra of the jet oscillations, shown in Fig.~\ref{fig:FFT}(b) as a function of the laser frequency for $P = 55$~mW (similar results were obtained for all laser powers higher than 11~mW). A strong dominance of the peak corresponding to the laser frequency can be observed in the range $f_\textrm{\small laser} \in [55, 120]$~Hz. Note that harmonics of $f_\textrm{\small laser}$ are also present in the spectrum of Fig.~\ref{fig:FFT}(b). In the range $f_\textrm{\small laser} < 55$~Hz the dominant peak corresponds to the first harmonic $2 f_\textrm{\small laser}$, and therefore in this region the jet's oscillations are also controlled by the laser induced perturbations.

\begin{figure}[h]
\centering
\includegraphics[width=8.6cm]{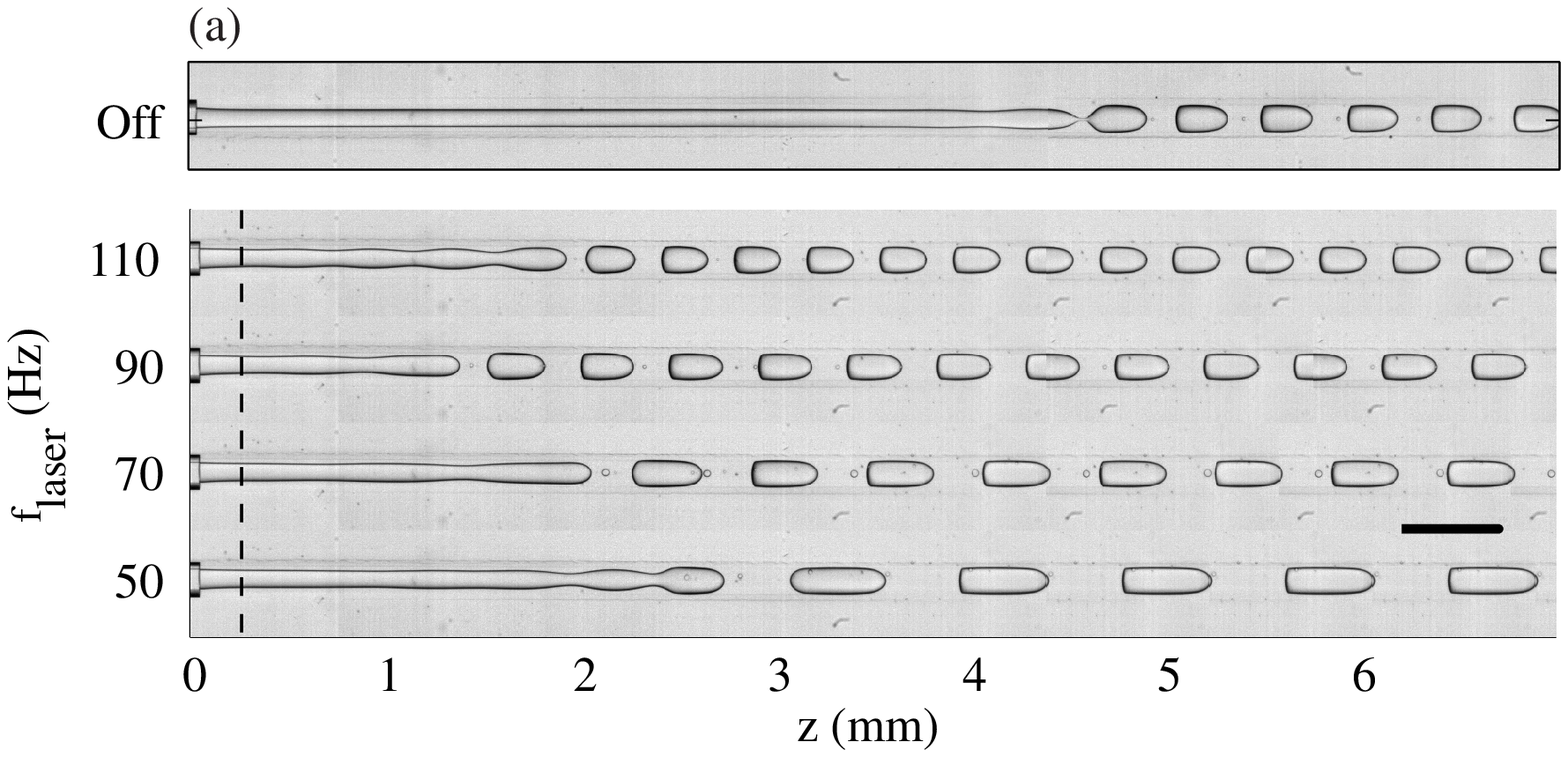}
\includegraphics{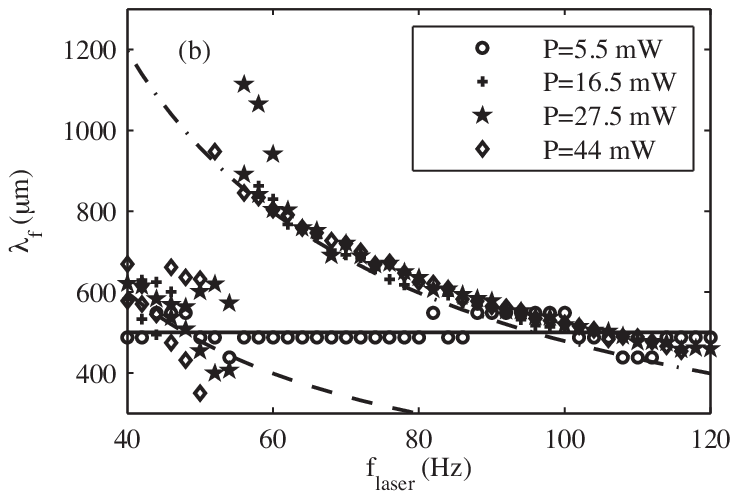}
\caption{\label{fig:Drops}(a) Images of the jet perturbed with the laser heating at a constant power $P = 55$~mW and different frequencies. The scale bar corresponds to 500~$\mu$m. The vertical dashed line marks the position of the laser focus. (b) Wavelength of the jet's radius oscillations as a function of the laser frequency $f_\textrm{\small laser}$ for different laser powers. The solid line corresponds to the dominant wavelength of the unperturbed jet's oscillations. The dot-dashed line corresponds to the curve $f_\textrm{\small laser} \lambda = v_0$, while the dashed line corresponds to $2f_\textrm{\small laser} \lambda = v_0$. Here, $Q_i = 20 \ \mu$L/min and $Q_e = 30 \ \mu$L/min.}
\end{figure}

Two main effects are observed in the frequency ranges where synchronization occurs: the length of the jet decreases as the laser power increases (not shown) and the drop size changes as a function of $f_\textrm{\small laser}$. Figure~\ref{fig:Drops}(a) shows snapshots of the jet forced by the laser with power $P = 55$~mW at various frequencies. One can observe that the jet shortens when the laser is turned on and that the drops are larger than without the laser forcing, decreasing with increasing $f_\textrm{\small laser}$.

By analysis of the drop size and considering the mean radius of the unperturbed jet, the wavelength of the forced jet oscillations $\lambda_f$ is computed using Eq.~(\ref{eq:DropSize}). Figure~\ref{fig:Drops}(b) shows $\lambda_f$ as a function of the laser frequency. In the range $f_\textrm{\small laser} \in [55, 120]$~Hz, the wavelength of oscillations decreases with the laser frequency, in close agreement with the relation $f_\textrm{\small laser} \lambda_f = v_0$. Here, $v_0$ is the phase velocity calculated from the dominant wavelength and frequency of the unforced state at the same flow rates. Note that for $f_\textrm{\small laser} < 55$~Hz, although the data are more scattered, the measured wavelengths are consistent with the relation $2 f_\textrm{\small laser} \lambda_f = v_0$. This again confirms the dominance of the first harmonic of $f_\textrm{\small laser}$ in the frequency spectrum.

\begin{figure}[h]
\centering
\includegraphics{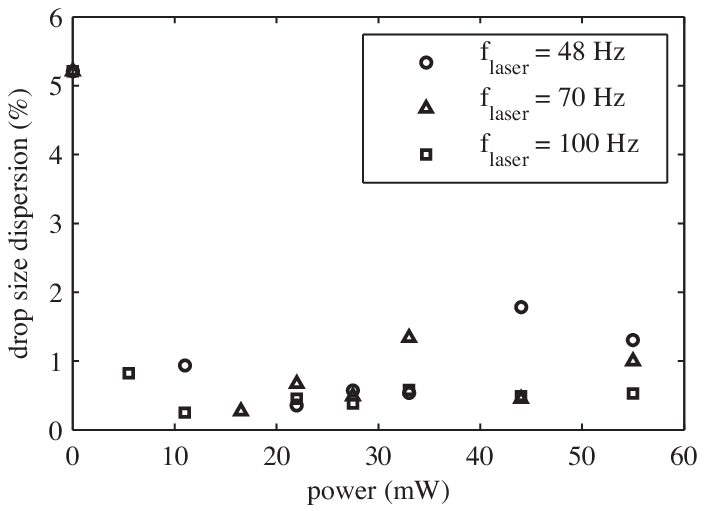}
\caption{\label{fig:Dispersion} Dispersion of the drop size distribution as a function of the laser frequency for the experimental condition $Q_i = 20 \ \mu$L/min and $Q_e = 30 \ \mu$L/min.}
\end{figure}

Finally the dispersion in the drop size, calculated as the standard deviation of the drop volume divided by the average, is shown in Fig.~\ref{fig:Dispersion} for three values of $f_\textrm{\small laser}$. Remarkably, the dispersion in the drop size, which is large in the jetting regime, decreases by a factor of about 5 in the presence of the laser forcing.

\section{Comparison between theory and experiments} \label{sec:Comparison}

Herrada et al.~\cite{Herrada2008} and Guillot et al.~\cite{Guillot2007, Guillot2008} have addressed the stability problem of the confined, infinitely long column of liquid shown in Fig.~\ref{fig:Jets}(a), when inertial effects are negligible. While Herrada et al. solve the problem analytically, Guillot et al. use, in addition, the lubrication approximation. Both theories, therefore, present similar results, although the latter is expected to be less accurate for perturbations with small wavelengths compared with the jet radius. In the parameter space explored in our study, however, this difference is irrelevant since the measured wavelengths are at least 10 times larger than the jet radius [Fig.~\ref{fig:BaseState}(c)].

The analyses of Herrada et al. and Guillot et al. yield a dispersion relation that relates the complex wave frequency $\omega$ to the complex wave number $k$ in the form:

\begin{equation}\label{eq:DispersionRelation}
\omega = v k + i \sigma(k),
\end{equation}

\noindent where $v$ is proportional to the velocity of the jet interface $U$, which depends on the flow rates and viscosities of the fluids. The growth rate $\sigma(k)$ is a nonlinear function of the wavenumber $k$, the fluid viscosities, flow rates and of the capillary number Ca. As a consequence of the nonlinearity the propagation of perturbations occurs in a dispersive way, meaning that perturbations with different wave numbers travel with different velocities along the jet surface. In particular, unstable perturbations travel with velocities ranging between $v^-$ and $v^+$, schematically shown in Fig.~\ref{fig:Jets}(b). The sign of $v^-$ determines if the flow is convectively or absolutely unstable. If $v^- < 0$, then there are unstable perturbations that travel upstream, making the flow absolutely unstable. On the contrary, if $v^- > 0$ all unstable perturbations travel downstream and the flow is convectively unstable~\cite{Monkewitz1990, Huerre1990}. Herrada et al. and Guillot et al. have found the transition between an absolutely and a convectively unstable flow (A/C transition) to depend on the capillary number Ca and the A/C transition in capillary jets has been traditionally related to the transition between dripping and jetting~\cite{Monkewitz1990}.

In the following, we test the predictive power of the theories by extracting several quantities that can be compared to our measurements. For that we note first that, contrary to the hypothesis of Herrada et al. and Guillot et al., our confining channel has square cross section instead of being cylindrical. As shown also by Guillot et al.~\cite{Guillot2008}, this mainly affects the theoretical results when the confinement of the jet $2R_i/h$ becomes close to one. We measure the confinement in our experiments in the range [0.38, 0.5] and therefore the geometry of our confinement channel can be considered, for calculation purposes, as a cylinder of equivalent radius $R_e = h/\sqrt{\pi} = 113 \ \mu$m.

When our experimental parameters ($\lambda = 0.6$, $\gamma = 40$~mN/m, $Q_e = 30 \ \mu$L/min and $R_e = 113 \ \mu$m) are injected into the theories considered here, the A/C transition is computed at $Q_i = 26.2 \ \mu$L/min with Guillot's theory and $Q_i = 23 \ \mu$L/min with Herrada's theory. These values are of the same order of magnitude than our measurement at $Q_i = 14 \ \mu$L/min. The disagreement in the exact value of the critical external flow rate could be related to the geometric differences in cross section between our square channel and the circular assumption in the theory.

Next, we compare the frequency behavior observed in the unforced flow regimes to the theoretical predictions. According to the spatiotemporal stability analysis, the behavior of absolutely and convectively unstable flows differs strongly. Absolutely unstable flows are dominated by a single, well defined frequency and therefore behave as self-sustained resonators pulsating at the absolute frequency $f_0$. This is because the perturbation with zero group velocity is unstable and, since it grows in place, it ultimately contaminates the whole domain~\cite{Huerre2000, Couairon1999, Chomaz2005}. In convectively unstable flows, conversely, instabilities are advected downstream and, without external noise, the flow would go back rapidly to the base state. Convectively unstable flows therefore amplify a broad range of unstable frequencies and the most amplified one, also noted for simplicity $f_0$, corresponds to the one with highest spatial growth rate~\cite{Huerre1990}.

\begin{figure}[h]
\centering
\includegraphics{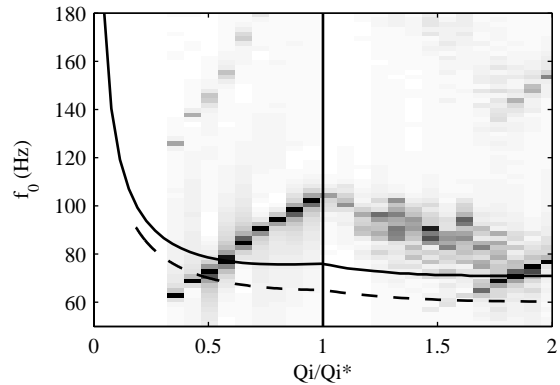}
\caption{\label{fig:GuillotHerrada} Frequency of the dominant perturbation predicted by  the theories of Guillot et al. (solid line) and Herrada et al. (dashed line) as a function of the inner flow rate normalized by the flow rate at which the A/C transition is predicted to occur. In this calculation, $Q_e = 30 \ \mu$L/min, $\eta_i = 60$~mPa~s, $\eta_e = 100$~mPa~s, $R_e = 113 \ \mu$m and $\gamma = 40$~mN/m. The underlying gray scale present our data of Fig.~\ref{fig:BaseState}(a) with $Q_i$ normalized by the observed transitional flow rate $Q_i=14 \ \mu$L/min.}
\end{figure}

With this in mind we use the dispersion relations from Eq. ~(\ref{eq:DispersionRelation}) to compute the frequency expected to dominate the flow, for both theories, as a function of the inner flow rate for an outer flow rate fixed at $Q_e = 30 \ \mu$L/min. In the absolute region the dominant frequency $f_0$ was calculated as the frequency of the perturbation with zero group velocity. In the convective region, $f_0$ was computed as the frequency of the mode with largest spatial growth rate $- k_{0,i}$, which is deduced from an inversion of the dispersion relation $\omega = \omega(k)$ to $k = k(\omega)$ with $\omega \in \Re$. The result is shown in Fig.~\ref{fig:GuillotHerrada} where the lines represent the theoretical predictions and the gray scale image recalls the experimental results of Fig.~\ref{fig:BaseState}(a). Both theories predict $f_0$ to exhibit a sharp decrease for small $Q_i$ and to grow slightly near the transition. In the convective region, $f_0$ decreases slowly after the transition and then grows again slowly for larger $Q_i$.

In the following sections we proceed to a detailed comparison between the theoretical predictions and our experimental observations in the different flow regimes.

\subsection{Dripping regime}

Our measurements of the flow in the dripping regime reveal most of the features of an absolutely unstable flow. Without external forcing, the dripping flow exhibits one well defined natural frequency that dominates the frequency spectrum, as shown in Figs.~\ref{fig:BaseState}(a) and (b). Moreover, when externally forced with the laser heating, the flow in the dripping regime does not respond to the forcing and remains dominated by the same natural frequency that dominates the unforced state [Fig.~\ref{fig:FFT}(a)]. However, this natural frequency does not appear to be linked to the absolute frequency predicted theoretically. Indeed, the theories predict $f_0$ to decrease steeply as a function of $Q_i$ for small $Q_i$ (Fig.~\ref{fig:GuillotHerrada}), which would result in the formation of larger drops at a smaller rate. On the contrary, we measure a linear increase in $f_0^\textrm{\small exp}$ with $Q_i$, which corresponds to the formation of drops of constant volume at a higher rate.

\subsection{First jetting regime}

In the first jetting regime, a broad range of frequencies are amplified, as one would expect if the flow is convectively unstable. Moreover, the J1 regime exhibits a decrease of $f_0^\textrm{\small exp}$ with the inner flow rate, which is coherent with the theoretical predictions in the convective regime near the A/C transition.

The hypothesis of a convectively unstable flow in the J1 regime is further confirmed by our experiments with the laser forcing. These experiments show the synchronization of the jet oscillations with the laser frequency, demonstrating the ability of the jet to amplify externally induced perturbations. Furthermore, a threshold laser power is necessary to synchronize the undulations of the jet to the laser frequency, but beyond this threshold the spectral response of the system is independent of the laser power [see Fig.~\ref{fig:Drops}(b)]. This is coherent with the linear picture proposed here where only the frequency of the external forcing is relevant to perturb the flow.

Finally, the amplified perturbations are found to follow the relation $f_\textrm{\small laser} \lambda_f = v_0$.  At leading order, this is coherent with the dispersion relation of  Eq.~(\ref{eq:DispersionRelation}).

\subsection{Second jetting regime}

\begin{figure}[h]
\centering
\includegraphics{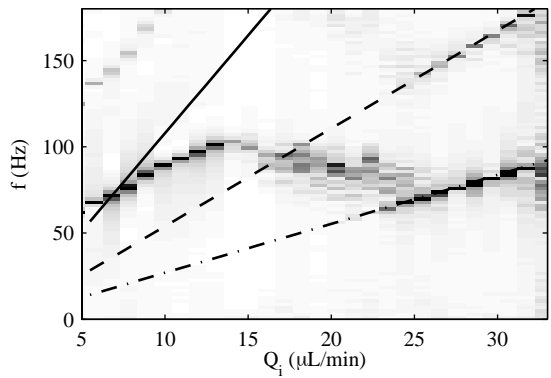}
\caption{\label{fig:SyringePump} Frequency of the perturbation induced by the injection of the inner fluid. The solid line represents the pumping rate [Eq.~(\ref{eq:PumpingRate})], while the dashed and dot-dashed lines show its second and fourth sub-harmonics. The latter is amplified by the flow.}
\end{figure}

In the second jetting regime a single, well defined peak dominates the frequency spectrum of the jet oscillations, which, at first, contradicts the hypothesis that the jetting regime corresponds to a convective instability. However the dominant frequency $f_0^\textrm{\small exp}$ increases linearly in the J2 regime, which allows us to relate this frequency to a perturbation associated with the syringe pump used to inject the inner fluid. In fact, the stepping frequency of the syringe pump increases linearly with $Q_i$ as

\begin{equation} \label{eq:PumpingRate}
f_{\rm pump} = \frac{4 Q_i}{\pi D^2 dx},
\end{equation}

\noindent where $D = 4.61$~mm is the syringe diameter and $dx = 0.088 \ \mu$m the syringe pump step. While $f_{\rm pump}$ is too high compared to $f_0^\textrm{\small exp}$ in the J2 regime, its fourth sub-harmonic ($f_{\rm pump}/4$) matches it precisely, as shown in Fig.~\ref{fig:SyringePump}. Note also that the second sub-harmonic ($f_{\rm pump}/2$) is also present in the frequency spectrum of the unforced state.

These results suggest that there is also a broad range of unstable frequencies in the J2 regime but the frequency imposed by the pumping rate of the inner fluid selects the dominant one. As can be seen from Fig.~\ref{fig:SyringePump}, the transition from the J1 regime with no dominant frequency to the J2 regime, with extrinsic frequency selection, occurs at the flow rate where the frequency imposed by the syringe pump enters the range of unstable frequencies. These results are coherent with a convective nature of the flow instability in the second jetting regime.

\section{Summary and conclusions} \label{sec:Conclusions}

The frequency response of the system can be summarized from Fig.~\ref{fig:BaseState}(a). In the dripping regime there is only one, well defined frequency which is selected intrinsically, although not directly linked to the absolute frequency of Refs.~\cite{Guillot2007, Herrada2008}. In the jetting regime, the zone of unstable frequencies is wider and with a less marked peak. The response of the J1 regime is typical of unselective noise amplification, with the whole range of unstable frequencies present in the spectrum, while the frequency selection in the J2 regime is extrinsic and is due to the frequency imposed by the syringe pump.

In conclusion, we have given experimental evidence to confirm that the jetting regime in a confined co-flowing system is related to a convective instability of the inner thread. Conversely, other mechanisms are likely involved in the dripping regime and determine the rate of drop formation. The frequency selection in this case could be influenced by a mechanism akin to the shear induced breakup described by Umbanhowar et al.~\cite{Umbanhowar2000}. This would suggest that the influence of the growth of hydrodynamic instabilities of confined co-axial jets in this regime is less important than the local effect of shear at the nozzle, for our particular geometry.

Confirmation of the instability nature of the flow in the jetting regime is important to understand the mechanisms that control drop formation when capillary and viscous effects are in competition. In particular, spatiotemporal stability analysis predicts high inherent dispersion in drop sizes in the jetting regime. This is because noise sources, such as micrometer sized fabrication defects of the channel walls or fluctuations in the pressure and flow rates, are unavoidably present. While this kind of perturbations are rapidly damped and usually negligible in microfluidics, they are exponentially amplified by a convectively unstable flow. Understanding the flow nature provides a way to control this polydispersity by inducing the amplification of a preferred single frequency with a controlled forcing. Here, we do so with periodic, localized laser heating, demonstrating the possibility of externally and dynamically tuning the frequency of drop formation and the drop size with low dispersion.

\acknowledgments

The authors acknowledge the help of Elise Lorenceau in the chip fabrication. MLC was partially funded by the EADS Corporate Foundation and by CONICYT Chile.


\begin{thebibliography}{10}

\bibitem{Eggers2008}
J.~Eggers and E.~Villermaux.
\newblock Physics of liquid jets.
\newblock {\em Rep. Prog. Phys.}, 71:036601, 2008.

\bibitem{Clanet1999}
C.~Clanet and J.~C. Lasheras.
\newblock Transition from dripping to jetting.
\newblock {\em J. Fluid Mech.}, 383:307--326, 1999.

\bibitem{Ambravaneswaran2000}
B.~Ambravaneswaran, S.~D. Phillips, and O.~A. Basaran.
\newblock Theoretical analysis of a dripping faucet.
\newblock {\em Phys. Rev. Lett.}, 85:5332--5335, 2000.

\bibitem{Coullet2005}
P.~Coullet, L.~Mahadevan, and C.~S. Riera.
\newblock Hydrodynamical models for the chaotic dripping faucet.
\newblock {\em J. Fluid Mech.}, 526:1--17, 2005.

\bibitem{Subramani2006}
H.~J. Subramani, H.~K. Yeoh, R.~Suryo, Q.~Xu, B.~Ambravaneswaran, and O.~A.
  Basaran.
\newblock Simplicity and complexity in a dripping faucet.
\newblock {\em Phys. Fluids}, 18:032106, 2006.

\bibitem{LeDizes1997}
S.~Le~Diz\`es.
\newblock Global modes in falling capillary jets.
\newblock {\em Eur. J. Mech. B}, 16:761--778, 1997.

\bibitem{Ganan-Calvo1998}
A.~M Ga\~n\'an{-}Calvo.
\newblock Generation of steady liquid microthreads and micron-sized
  monodisperse sprays in gas streams.
\newblock {\em Phys. Rev. Lett.}, 80:285--288, 1998.

\bibitem{Baroud2010}
C.~N. Baroud, F.~Gallaire, and R.~Dangla.
\newblock Dynamics of microfluidic droplets.
\newblock {\em Lab Chip}, 10:2032--2045, 2010.

\bibitem{Utada2007a}
A.~S. Utada, A.~Fern\'andez-Nieves, H.~A. Stone, and D.~A. Weitz.
\newblock Dripping to jetting transitions in coflowing liquid streams.
\newblock {\em Phys. Rev. Lett.}, 99:094502, 2007.

\bibitem{Guillot2007}
P.~Guillot, A.~Colin, A.~S. Utada, and A.~Ajdari.
\newblock Stability of a jet in confined pressure-driven biphasic flows at low
  {Re}ynolds numbers.
\newblock {\em Phys. Rev. Lett.}, 99:204502, 2007.

\bibitem{Guillot2008}
P.~Guillot, A.~Colin, and A.~Ajdari.
\newblock Stability of a jet in confined pressure-driven biphasic flows at low
  {R}eynolds number in various geometries.
\newblock {\em Phys. Rev. E}, 78:016307, 2008.

\bibitem{Rayleigh1879a}
Lord Rayleigh.
\newblock On the stability, or instability, of certain fluid motion.
\newblock {\em Proc. Lond. Math. Soc.}, 10:4, 1879.

\bibitem{Leib1986b}
S.~J. Leib and M.~E. Goldstein.
\newblock Convective and absolute instability of a viscous liquid jet.
\newblock {\em Phys. Fluids}, 29:952--954, 1986.

\bibitem{Monkewitz1990}
P.~A. Monkewitz.
\newblock The role of absolute and convective instability in predicting the
  behavior of fluid systems.
\newblock {\em Eur. J. Mech., B/Fluids}, 9:395--413, 1990.

\bibitem{Herrada2008}
M.~A. Herrada, A.~M. Ga\~n\'an{-}Calvo, and P.~Guillot.
\newblock Spatiotemporal instability of a confined capillary jet.
\newblock {\em Phys. Rev. E}, 78:046312, 2008.

\bibitem{Ody2007}
C.~P. Ody, C.~N. Baroud, and E.~de~Langre.
\newblock Transport of wetting liquid plugs in bifurcating microfluidic
  channels.
\newblock {\em J. Coll. Inter. Sc.}, 308:231--238, 2007.

\bibitem{Handbook}
D.~R. Lide, editor.
\newblock {\em CRC Handbook of chemistry and physics}.
\newblock CRC Press LLC, 2003.

\bibitem{Cordero2009a}
M.~L. Cordero, E.~Verneuil, F.~Gallaire, and C.~N. Baroud.
\newblock Time-resolved temperature rise in a thin liquid film due to laser
  absorption.
\newblock {\em Phys. Rev. E}, 79:011201, 2009.

\bibitem{Baroud2007a}
C.~N. Baroud, J.~P. Delville, F.~Gallaire, and R.~Wunenburger.
\newblock Thermocapillary valve for droplet production and sorting.
\newblock {\em Phys. Rev. E}, 75:046302, 2007.

\bibitem{Cubaud2008}
T.~Cubaud and T.~G. Mason.
\newblock Capillary threads and viscous droplets in square microchannels.
\newblock {\em Phys. Fluids}, 20:053302, 2008.

\bibitem{Huerre1990}
P.~Huerre and P.~A. Monkewitz.
\newblock Local and global instabilities in spatially developing flows.
\newblock {\em Ann. Rev. Fluid Mech.}, 22:473--537, 1990.

\bibitem{Huerre2000}
P.~Huerre.
\newblock Open shear flow instabilities.
\newblock In G.K. Batchelor, H.K. Moffatt, and M.G. Worster, editors, {\em
  Perpectives in Fluid Dynamics}, pages 159--229. Cambridge University Press,
  2000.

\bibitem{Couairon1999}
A.~Couairon and J.-M. Chomaz.
\newblock Fully nonlinear global modes in slowly varying flows.
\newblock {\em Phys. Fluids}, 11:3688--3703, 1999.

\bibitem{Chomaz2005}
J.~M. Chomaz.
\newblock Global instabilities in spatially developing flows: Non-normality and
  nonlinearity.
\newblock {\em Ann. Rev. Fluid Mech.}, 37:357--392, 2005.

\bibitem{Umbanhowar2000}
P.~B. Umbanhowar, V.~Prasad, and D.~A. Weitz.
\newblock Monodisperse emulsion generation via droplet break off in a coflowing
  stream.
\newblock {\em Langmuir}, 16:347--351, 2000.

\end{thebibliography}

\end{document}